\documentclass{emulateapj}                   

\usepackage{amsmath}
\usepackage{amssymb}

\newcommand{\kms}{\ensuremath{\mathrm{km~s}^{-1}}}
\newcommand{\ergss}{\ensuremath{\mathrm{ergs~s}^{-1}}}
\newcommand{\msun}{\ensuremath{M_\odot}}
\newcommand{\xb}{\ensuremath{\bar{x}}}

\newcommand{\fth}{\ensuremath{\frac{3}{2}}}
\newcommand{\ffh}{\ensuremath{\frac{5}{2}}}
\newcommand{\eT}{\ensuremath{\frac{\epsilon_i}{kT}}}

\newcommand{\Ms}{\ensuremath{{M_\star}}}
\newcommand{\ms}{\ensuremath{{m_\star}}}
\newcommand{\rs}{\ensuremath{{r_\star}}}

\newcommand{\Rs}{\ensuremath{{R_\star}}}
\newcommand{\ps}{\ensuremath{{\rho_\star}}}
\newcommand{\Ts}{\ensuremath{{T_\star}}}
\newcommand{\gcm}{\ensuremath{{\rm g~cm}^3}}
\newcommand{\gop}{\ensuremath{{\rm cm}^2~{\rm g}^{-1}}}
\newcommand{\Eint}{\ensuremath{{E_{\rm int}}}}
\newcommand{\EKE}{\ensuremath{{E_{\rm KE}}}}
\newcommand{\km}{\ensuremath{{\kappa_{\rm min}}}}
\newcommand{\Mbh}{\ensuremath{M_{\rm bh}}}
\newcommand{\tfb}{\ensuremath{t_{\rm fb}}}

\shortauthors{Kasen, D., Ramirez-Ruiz, E.}
\shorttitle{Tidal Disruption Recombination Transients}

\begin{document}

\title{Optical Transients from the Unbound Debris of Tidal Disruption}

\author{Daniel Kasen\altaffilmark{1,2}\email{kasen@ucolick.org} Enrico Ramirez-Ruiz\altaffilmark{1}}

\altaffiltext{1}{University of California, Santa Cruz}
\altaffiltext{2}{Hubble Fellow}

\begin{abstract} 

  In the tidal disruption of a star by a black hole, roughly
  half of the stellar mass becomes bound and falls into the black
  hole, while the other half is ejected at high velocity.  Several
  previous studies have considered the emission resulting from the
  accretion of bound material; we consider the possibility that the
  unbound debris may also radiate once it has expanded and become
  transparent.  We show that the gradual energy input from hydrogen
  recombination compensates for adiabatic loses over significant
  expansion factors.  The opacity also drops dramatically with
  recombination, and the internal energy can be radiated by means of a
  cooling-transparency wave propagating from the surface layers
  inward.  The result is a brief optical transient occurring $\sim 1$
  week after disruption and lasting 3-5 days with peak luminosities of
  $10^{40}-10^{42}$~\ergss, depending on the mass of the disrupted
  star.  These recombination powered transients should accompany the
  x-ray/ultraviolet flare from the accretion of bound material, and so
  may be a useful signature for discriminating tidal disruption
  events, especially for lower and intermediate mass black holes.
\end{abstract}


\section{Introduction}
Stars unfortunate enough to pass too close to a massive black hole are
torn apart by tidal gravity forces.  In the process, roughly half of
the stellar material becomes bound, circularizes, and may eventually
accrete onto the black-hole.  The other half of the star is ejected
from the system at high differential velocities.

Tidal disruption events offers one means of probing a central black
hole in quiescent galaxies.  The dominant observational signature
should be a bright x-ray flare powered by accretion of bound
material onto the black hole
\citep{Rees_1988,Ulmer_1999,Ramirez_dis09}.  Candidate disruption
flares have been detected in x-ray/ultraviolet surveys
\citep{Donley_Flare, Gezari_2006, Esquej_Flare} in some case with
counterpart emission in the optical \citep{Gezari_2009}.  The light
curves decline as power laws with luminosities comparable to the
Eddington luminosity for $10^6-10^7$~\msun\ black holes.
Nevertheless, it can be difficult to uniquely identify these flares as
tidal disruption events as opposed to other forms of nuclear activity.

It is therefore useful to determine additional discriminating
signatures of tidal disruption events. A few possibilities have been
suggested.  For deeply penetrating orbits, strong tidal compression
perpendicular to the orbital plane leads to the formation of shocks,
which breakout out in short ($\sim 100$~sec) but bright ($L \sim
10^{43}$~\ergss) x-ray bursts \citep{Kobayashi_TD, Brassart_2008,
  Guillochon_TD}. Later on, streams of material falling onto the black
hole may undergo collisions producing short UV flares of luminosity
$10^{40}-10^{41}$~\ergss\ \citep{Cannizzo_1990,Kim_TD}.  In addition,
the x-ray/UV luminosity from accretion may be absorbed and reprocessed
by the unbound debris into optical emission, leading to optical luminosities
of $\sim 10^{40}$~\ergss\ \citep{Bogdi_TD, Strubbe_TD}. If a
substantial fraction of the accreting material is blown in a wind, 
the optical radiation escaping from this outflow may be much brighter,
$10^{42}-10^{43}$~\ergss\ \citep{Strubbe_TD}.  Optical emission is of
particular interest, as observational surveys are typically most
sensitive at these wavelength.

Little attention has been given to  emission from the unbound
debris of disruption.  This material, which expands differentially
with velocities $\sim 10^4$~\kms\ in some ways resembles a supernova
remnant, though usually lacking any radioactive isotopes to power the
light curve.  While the initial internal energy of the star can be
fairly significant ($\Eint \sim 10^{48}$~ergs), most of it will be
lost to adiabatic expansion before the ejecta becomes sufficiently
transparent to radiate.  For an ideal gas with adiabatic index $\gamma
= 5/3$, the energy and temperature evolve adiabatically as $E \propto
\rho^{(\gamma -1)} \propto \rho^{2/3}$.  A decline in density by at
least a factor of $\sim 10^9$ is required before the debris might
become translucent, at which point the internal energy and temperature
will have decreased by a factor $\sim 10^6$.  Hence the apparent lack
of interest in diffusive thermal emission from the unbound remnants of tidal
disruption.

There is, however, another important energy reservoir of the debris --
ionization energy, which amounts to $\sim 10^{46}$~ergs per solar mass
of ionized hydrogen and is not subject to adiabatic degradation.  When
recombination eventually sets in at temperatures $T_i \sim 10^4$~K,
the energy released per atom $\epsilon_i$ is an order of magnitude
greater than the mean thermal energy, $\epsilon_i/kT_i \approx 15$.
Recombination is therefore a gradual process, steadily inputting
energy to maintain the gas temperature near $T_i$. The thermodynamic
evolution is described by $E \propto \rho^{(\gamma_3 -1)}$ where
$\gamma_3$ is the third generalized adiabatic index, now a function of
density and temperature.  In the region of partial ionization
$\gamma_3 - 1 \approx 2 kT/\epsilon_i \approx 0.1$ (see
\S\ref{sec:thermo}) so that the energy decline is indeed gradual.

When the effects of ionization are included, the debris can expand by
a much larger factor, and reach much lower densities, before the
internal energy is depleted.  When near neutrality is reached, the
opacity also drops sharply due to the elimination of electron
scattering.  It may be possible for radiation to escape at this time,
taking the form of a cooling-transparency wave that propagates from
the surface inwards.  The escape of photons will be aided by the
aspherical geometry of the debris, which has been stretched by tidal
forces into a thin stream, with lower optical depth along the
direction perpendicular to the orbital plane

It is therefore plausible that the unbound debris of tidal disruptions
gives rise to a brief optical transient powered primarily by the
energy tapped from recombination.  We call such an event a
recombination transient (RT). The light curves, even if they are dim,
might be relevant for upcoming synoptic optical surveys which will
probe for the first time extra-galactic transients with luminosities
between classical novae ($M_R > -10$~mag) and supernovae ($M_R <
-14$~mag).  Here we elaborate on the basic physical ideas and make
approximate predictions of the light curves.

\section{The Debris of Tidal Disruption}

\begin{figure}
\includegraphics[width=3.5in]{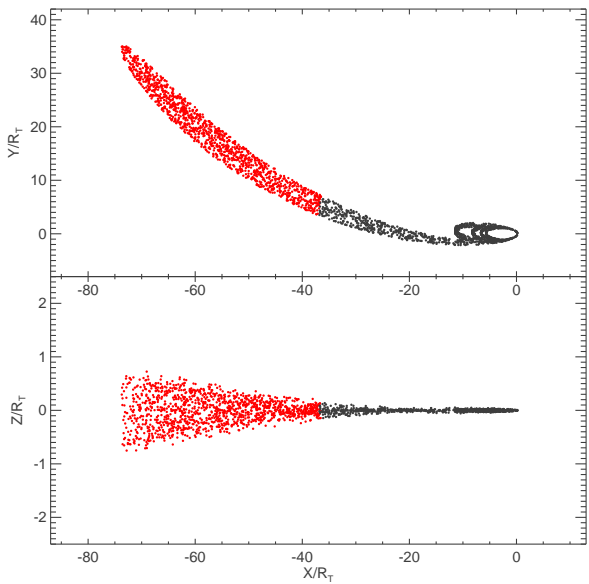}
\caption{Geometrical distribution of the debris (2 days after pericenter
passage) for a $\beta = 5$
 tidal disruption of a solar type star by a $10^6$~\msun\ black hole.
 The location of individual points was determined by following the collisionless orbits
of particles under the gravity of the black hole, starting from a pancake configuration at pericenter. The top panel
 shows a slice through the orbital plane, while the bottom panel
 shows the distribution perpendicular to the orbital plane (note the
 vertical axis of the bottom panel is zoomed by a factor of 10).
  Black points denote bound material and red points unbound material.
  \label{Fig:debris}}
\end{figure}

The dynamics of tidal disruption have been widely studied, both
analytically and numerically \citep[e.g.,][]{Rees_1988, CL_83,
  Kochanek_TD, Brassart_2008, Lodato_TD,Guillochon_TD,Ramirez_dis09}.
Disruption occurs when a star passes within the tidal radius $R_T =
\Rs (\Mbh/\Ms)^{1/3}$, where \Mbh\ is the black hole mass, and \Ms\ and
\Rs\ the stellar mass and radius.  The penetration factor $\beta =
R_T/R_p$ describes the depth of the encounter, where $R_p$ is the
radius at pericenter.  The velocity at pericenter is $v_p = (2 G
\Mbh/R_p)^{1/2}$ or
\begin{equation}
\begin{split}
v_p &\approx 6 \times 10^4~\beta^{1/2}~m_6^{1/3} \ms^{1/6}
\rs^{-1/2}~\kms
\end{split}
\end{equation}
where $m_6 = \Mbh/(10^6 \msun)$, $\ms = \Ms/M_\odot$, and $\rs =
\Rs/R_\odot$.  Because the initial velocity of the star at infinity
($\sim 100$~\kms) is much less than $v_p$, it is assumed to approach
on an essentially parabolic orbit.

Within $R_T$, the self-gravity of the star becomes sub-dominant.
Material on the far side of the star, being more distant to to the
black hole by \Rs, possesses an excess specific energy relative to the
center of mass, $\Delta \epsilon = G \Mbh \Rs /R_p^2$.  This material
becomes unbound with an expansion velocity at infinity $v_\infty = (2
\Delta \epsilon)^{1/2}$ or
\begin{equation} 
v_\infty \approx 6 \times 10^3 ~\beta~m_6^{1/6} \ms^{1/3} \rs^{-1/2}~\kms
\end{equation}
The half of the star nearer to the black hole, with specific energy
$-\Delta \epsilon$, becomes bound and eventually may accrete onto the
black hole.

The unbound material is ejected from the system on hyperbolic orbits,
and the debris elongates into a thin stream
\citep{khokhlov_disrupt,Kochanek_TD,rosswog09}.  We describe the final
geometry with the parameters $R_o, R_t$, and $R_z$ which denote,
respectively, the spatial extant of the debris in the orbital
direction, the transverse direction within the orbital plane, and the
direction perpendicular to the orbital plane. The thickness of the
stream within the orbital plane is related to the difference in
specific energy between the leading and trailing edges of the star as
it approaches pericenter, $\Delta \epsilon_t = G \Mbh \Rs^2/R_p^3$.
Comparing the corresponding velocity at infinity $v_t = (2 \Delta
\epsilon_t)^{1/2}$ to $v_\infty$
 provides a estimate of the axis ratio
$E_t = R_o/R_t$.
\begin{equation} E_t \approx v_\infty/v_t
\approx 10~\beta^{-1/2} (m_6/\ms)^{1/6}
\end{equation}

In the dimension perpendicular to the orbital plane (the
$z$-direction) tidal forces compress the star into a pancake
\citep{Guillochon_TD}. The compression raises the central density by a
factor proportional to $\beta^3$ \citep{CL_83}, with a corresponding
adiabatic increase in temperature of $T \propto \rho^{(\gamma-1)}
\propto \beta^2$. Eventually the star rebounds under the increased
central pressure.  For deep penetrations ($\beta \ga 3$), shocks form
and propagate outward in the $z$-direction.  The temperature of the
shocked material will be of order the virial temperature
\begin{equation}
\Ts \approx \beta^2 \frac{ G \Ms \mu m_p}{\Rs k_b} \approx 1 \times
10^7 ~\beta^2 \ms \rs^{-1}~{\rm K}
\end{equation}
where $\mu = 0.5$ is the mean particle mass of ionized hydrogen.  The
velocity in the z-direction of the shocked material will be of order
\begin{equation}
v_z \approx  (3 k  \Ts/\mu m_p)^{1/2}
\approx 600~\beta~\ms^{1/2} \rs^{-1/2}~\kms
\label{Eq:vz}
\end{equation}
Comparable shock velocities can be seen in the hydrodynamical
calculations of \cite{Brassart_2008}.  This z-velocity serves to
increase final the spatial extent of the unbound debris in the
$z$-direction.  The orbital plane of a gas element with non-zero
z-velocity at pericenter will be inclined to the plane of the stellar
orbit by an angle $\arcsin (v_z/v_p)$.  The axis ratio of the debris
in the direction perpendicular to the orbital plane $E_z = R_o/R_z$ is
then approximately
\begin{equation}
E_z \approx v_p/v_z = 100~\beta^{-1/2} (m_6/\ms)^{1/3} 
\end{equation}

We can model approximately the final shape of the unbound debris by
simply following the orbits of collisionless test particles in the
gravitational potential of the black hole, while ignoring
self-gravity.  For initial conditions at pericenter we take a uniform
pancake of radius $R_\odot$ moving in the y-direction with speed $v_p$
and with a $z$-component of velocity given by Eq.~\ref{Eq:vz}.  Two
days later, the final thin stream geometry (Figure~\ref{Fig:debris})
is roughly described by axis ratios of $E_z = 100$ and $E_t = 10$.

The unbound debris eventually enters a homologous phase and the volume
scales as $V(t) = \Rs^3 \zeta^3$ where we define the expansion
parameter:
\begin{equation}
\zeta = (R_o R_t R_z)^{1/3} \frac{1}{\Rs} = \frac{v_\infty t }{\Rs}
(E_z E_t)^{-1/3}
\end{equation}
We can also write $\zeta(t) = t/t_e$ where the expansion time is
defined
\begin{equation}
\begin{split}
t_e &= \frac{\Rs}{v_\infty} (E_z E_t)^{1/3}  \\
&= 1120
~\beta^{-1} m_6^{-1/6} \ms^{-1/3} \rs^{3/2}
\biggl(\frac{ E_z E_t}{1000} \biggr)^{1/3} 
~{\rm sec}
\end{split}
\end{equation}

Initially, the stellar mass and energy are centrally concentrated, but
shocks largely homogenize the density and temperature structures in
the vertical direction \citep{Brassart_2008,Guillochon_TD,rosswog09}.
This outward transport of entropy is an important factor in obtaining
a bright transient.  The total internal energy of the unbound debris
after it has rebounded to its original volume will be of order its
original value
\begin{equation}
\Eint = f_E \frac{1}{2}\frac{G \Ms^2}{\Rs} \approx  
2\times 10^{48}~f_E \ms^2 \rs^{-1}
~{\rm ergs}.
\label{Eq:init_energy}
\end{equation}
where $f_E$ is a constant of order unity.  By comparison, the total
kinetic energy at infinity of the unbound debris is of order $E_{\rm
  KE} = 1/4 \Ms v_\infty^2 \approx \beta^2 2 \times 10^{50}$~ergs.
The ratio of internal to kinetic energy is initially $\Eint/\EKE
\approx 10^{-2} \beta^{-2} \ll 1$, which can be contrasted with that
of shock driven Type~II supernovae, $\Eint/\EKE \approx 1$. The
difference in initial entropy results in a distinct thermodynamic
evolution and light curve for the two events.

\section{Thermodynamic Evolution}
\label{sec:thermo}

\begin{figure}
\includegraphics[width=3.5in]{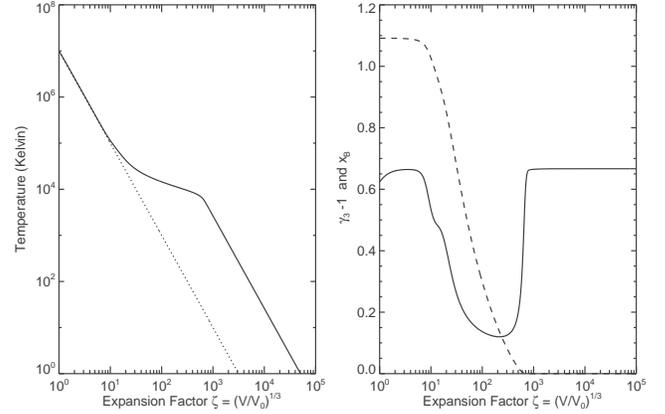}
\caption{Thermodynamic evolution of a parcel of gas with initial
  temperature $T_0 = 2\times 10^7$~K and density $\rho_0 = 1$~\gcm.
  {\it Left:} Temperature (solid line) as a function of expansion
  factor in radius.  The dashed line shows the $T \propto \zeta^{-2}$
  law for a constant $\gamma = 5/3$ gas. The plateau in temperature
  starting at $\zeta \approx 10$ is due to the energy input from
  recombination.  {\it Right:} Evolution of the effective adiabatic
  index $\gamma_3 -1$ (solid line) and the mean ionization fraction
  \xb\ (dotted line) as a function of the expansion factor.  In the
  region of partial ionization, $\gamma_3 -1 \approx 0.1$. \\
\label{Fig:T_evolve}}
\end{figure}

Following disruption, the evolution of the expanding debris follows
the thermodynamics of ionizing gases
\citep{Krishna_Swamy,Mihalas_MIhalas}.  Initially the material is
extremely optically thick and the evolution essentially adiabatic, $dE
= -P dV$, where the pressure is
\begin{equation}
P = (1 + \xb) \frac{N k T}{V} + \frac{1}{3}a T^4,
\label{Eq:P}
\end{equation}
with $N$ the number of nuclei and $V$ the volume.  The mean ionization
state is $\xb = \sum_i A_i x_i$, where $x_i$ is the ionization
fraction and $A_i$ the abundance fraction of species $i$.  For our
purposes, only hydrogen and two ionization states of helium need be
considered. The total internal energy is
\begin{equation}
E = \fth(1 + \xb) N k T +  a T^4 V + \sum_i N A_i x_i \epsilon_i 
\label{Eq:energy}
\end{equation}
Excitation energy terms can be ignored.  Once the temperature is high
enough to excite the atoms the gas will quickly become ionized.

Assuming local thermodynamic equilibrium the ionization state of each
species is given by the Saha equation:
\begin{equation}
\frac{x_i}{1 - x_i}  \frac{\xb}{V} = \frac{c_i}{Z_i} T^{3/2} \exp{-\eT}
\label{Eq:saha}
\end{equation}
where $c_i$ is a constant, and $Z_i$ is the partition function, which
we take to be independent of temperature and density.

From Eq.~\ref{Eq:energy} we have
\begin{equation}
\begin{split}
dE = 
N k T  \sum A_i dx_i \biggl[  \fth + \eT \biggr]
+
\\ \frac{dT}{T} \biggl[ \fth (1 + \xb) N k T + 4 a T^4 V \biggr] 
+ \frac{dV}{V} a T^4 V
\label{Eq:dE}
\end{split}
\end{equation}
while the logarithmic derivative of Eq.~\ref{Eq:saha} gives:
\begin{equation}
\frac{d x_i}{x_i(1 - x_i)} 
+ \frac{d \xb}{\xb}
- \frac{dV}{V} =
 \frac{dT}{T} \biggl[ \ffh + \eT \biggr]
\label{Eq:dx}
\end{equation}
Because the ionization energies of hydrogen and helium are fairly well
separated, we approximate $d \xb = dx_i$

Combining Eqs.~\ref{Eq:dE}, \ref{Eq:dx}, and \ref{Eq:P} gives for
adiabatic evolution
\begin{equation}
\frac{dT}{T} = (\gamma_3 - 1) \frac{dV}{V}
\label{Eq:Tevolve}
\end{equation}
where
\begin{equation}
\gamma_3 = 1 + \frac
{ (1 + \xb)(1 + 4 \alpha) +  \sum_i A_i \frac{x(1-x)}{(2-x)} (\fth + \eT) }
{ (1 + \xb)( \fth+ 12 \alpha) + \sum_i A_i \frac{x(1-x)}{(2-x)} (\fth + \eT)^2}
\label{Eq:g3}
\end{equation}
were $\alpha = aT^4 V/ 3 N k T (1 + \xb)$ is the ratio of radiation
pressure to gas pressure. The sum here runs over both ionization
states of helium, treated independently.

When either the gas or radiation terms dominate, Eq.~\ref{Eq:g3}
reduces as expected to $\gamma_3 = 5/3$ or $4/3$, respectively.  The
terms under the sums, which incorporate the effects of ionization, are
zero for $x = 0$ and $1$ but maximal in the region of partial
ionization of hydrogen.  For $x_H \approx 0.6$ we get the smallest
values of $\gamma_3$
\begin{equation}
(\gamma_3 - 1)_{\rm min} \approx \frac{k_B T}{\epsilon_H} 
\biggl[1 + 10 \frac{k_b T}{\epsilon_H} (1 +\alpha) \biggr]
\approx 0.1~\biggl(\frac{T}{10^4~{\rm K}}\biggr)
\label{Eq:g3min}
\end{equation}
where we used the fact that $\epsilon_H/k_B T \approx 15$ for
temperatures $T \approx 10^4$~K.

From Eq~\ref{Eq:g3min} we can estimate the amount of expansion the
debris will undergo before recombination is complete.  For an initial
temperature of $T = 10^7$~K and density $\rho = 1$~\gcm, gas energy
dominates and one has at first $T \propto \zeta^{-2}$.  The debris
will expand by a factor $\zeta \approx 10$ before the temperature
drops into the regime of recombination at $T \sim 4 \times
10^4$~K. Hydrogen remains partially ionized for only a narrow
temperature range, however this occurs over an extended period of time
-- to reduce $T$ from $4\times 10^4$~K down to $1 \times 10^4$
requires additional expansion by a factor $\sim
4^{1/3(\gamma_3-1)_{\rm min}} \sim 100$.  Thus, for these initial
conditions, we anticipate total expansion of $\zeta_R \approx 1000$
before recombination is complete.

Figure~\ref{Fig:T_evolve} shows the thermodynamic evolution of a
parcel of gas with the above initial conditions, as determined from a
direct numerical integration of Eq.~\ref{Eq:Tevolve}.  The initial $T
\propto \zeta^{-2}$ law plateaus at $\zeta \approx 10$, after which
the gradual input of energy from recombination maintains the
temperature at a near constant value.  As expected, recombination is
essentially complete by $\zeta_R \approx 1000$, and the $T \propto
\zeta^{-2}$ behavior resumes.  Naturally, a higher initial temperature
allows for larger expansion factors, but so does a lower density.  
This is because at lower density
recombination occurs at a lower temperature, which implies a lower
$\gamma_3-1$ in the recombination regime (Eq.~\ref{Eq:g3min}).

\section{Opacity and Radiative Transfer}

\begin{figure}
\includegraphics[width=3.5in]{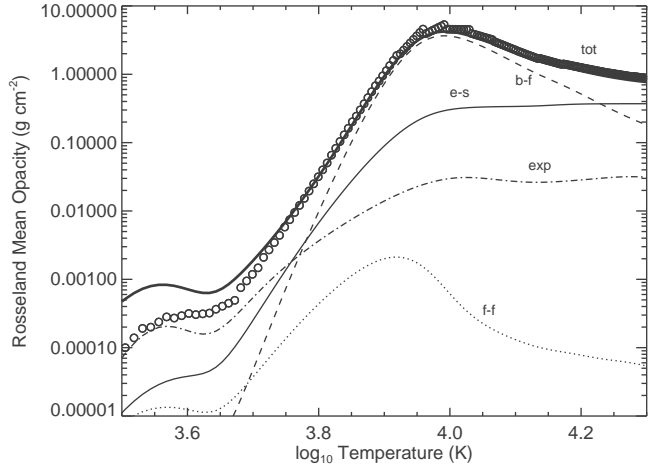}
\caption{Rosseland mean opacity as a function of temperature for solar
  metallicity material with density $\rho = 10^{-10}$ at a time $t =
  5\times 10^5$~sec. The thick solid line shows the total opacity, which is
  the sum of the opacity from bound-free (dashed line), electron scattering (thin solid line),
  free-free (dotted line) and line expansion opacity (dot-dashed).  The open circles are from the tabulation of the Opacity
  Project.
\label{Fig:opac}}
\end{figure}

After the debris has expanded by a factor $\zeta_R \sim 1000$ and
recombined, its internal energy is around $10^{45}$~ergs per \msun\ of
hydrogen.  This is $\sim 500$ times greater than what is predicted
for a constant $\gamma = 5/3$ evolution, demonstrating the critical
importance of the ionization effects.  If the energy can be radiated
on the time scale of days, a bright transient may be produced.
However, if the debris remains optically thick the energy will lost
quickly to expansion in the gas dominated regime.

Radiation will preferentially escape along the z-direction, the
shortest dimension with extant $R_z = \Rs \zeta E_z^{-2/3}
E_t^{1/3}$. The total optical depth along this direction is
$\tau_z = \rho \kappa_R R_z$, or taking an average value for the density
\begin{equation}
\tau_z
= 7 \times 10^{3}~\kappa_R \zeta_3^{-2} \ms \rs^{-2} 
\biggl( \frac{100}{E_z} \biggr)^{2/3} 
\biggl( \frac{E_t}{10} \biggr)^{1/3}
\label{Eq:tau}
\end{equation}
where $\zeta_3 =
\zeta/10^3$ and $\kappa_R$ is the Rosseland mean opacity in units
of cm$^{2}$~g$^{-1}$.  The debris
is very optically thick for $\kappa_{\rm es}
\approx 0.4$~\gop, the value for Thomson scattering in ionized
hydrogen. However, the opacity drops dramatically with hydrogen
recombination.  From Eq.~\ref{Eq:saha}, the free electron density
decreases by a factor $\sim \exp(-\epsilon_H/2 k_B T)
\sim 10^{-3}$ as the temperature changes from $10^4$~K to $5000$~K.

Other relevant sources of opacity likewise decline sharply for $T <
10^4$; free-free because of the declining electron density and
bound-free because of the reduced occupation number of excited levels.
The $H^-$ and molecular opacities, however, may start to become more
significant at lower temperatures.  In addition, Doppler broadening in
the differentially expanding debris enhances the line opacity from
numerous iron group lines.  In Figure~\ref{Fig:opac} we plot the 
various sources of opacity as a function of temperature.  The opacity
reaches a minimum value of $\km \approx  5 \times 10^{-4}$~\gop\ at $T
\approx 5000$~K.

The unbound debris may thus begin to radiate effectively when the
outermost layers of material have cooled to the ``transparency''
temperature $T_t \approx 5000$~K.  This emission should begin at
roughly a time $t_t = \zeta_R t_e$ after pericenter passage
\begin{equation}
t_t = 13~
~\zeta_3 \beta^{-1} m_6^{-1/6} \ms^{-1/3} \rs^{3/2}
\biggl(\frac{ E_z E_t}{1000} \biggr)^{1/3} 
~{\rm days}
\label{Eq:RT_t}
\end{equation}
Around this time, a sharp recombination front develops in the outer
layers of debris -- ionized material below the front is optically
thick, while neutral material above will be optically thin if the
densities are sufficiently low.  Radiative cooling at the front
promotes further recombination, causing this surface to move
progressively inward in mass coordinates.  The so-called transparency
wave \citep{Zeldovich_Raizer, Grassberg_1976} allows for a more rapid
release of the internal energy.

The condition for a transparency wave to successfully propagate is
that the optical depth evaluated at the minimum opacity be of order
unity.  From Eq.~\ref{Eq:tau} and $\km = 5\times 10^{-4}$ this occurs
for expansion factors $\zeta \ga 1500$.  The thermodynamic evolution
of \S\ref{sec:thermo} predicts expansion factors of $\zeta_R
\sim 1000$ before recombination is complete.  The coincidental near
equality of these numbers indicates that it is possible, though just marginally so, for a
recombination wave to occur and radiate away a significant fraction of
the remaining internal energy. 

The photosphere of the debris, which is coincident with the
recombination front, cools by radiating a flux $\sigma T_t^4$.
Assuming that gas interior to the front is nearly neutral at a
temperature near $T_t$, the speed at which transparency wave releases
the internal energy and propagates inward is
\begin{equation}
v_{\rm tw} = \frac{\sigma T_t^4}{3/2 n k T_t}
\approx 600~\zeta_{3}^3 \ms^{-1} \rs^3 
\biggl( \frac{T_t}{5000~{\rm K}} \biggr)^3
~\kms
\end{equation}
The timescale for the light curve to decline after peak will be
comparable to the time it takes the transparency wave to propagate the
distance $R_z$
\begin{equation}
t_{\rm tw} =  1.3~\zeta_{3}^{-2} \ms \rs^{-2}
\biggl( \frac{100}{E_z} \biggr)^{2/3} 
\biggl( \frac{E_t}{10} \biggr)^{1/3}
\biggl( \frac{T_t}{5000~{\rm K}} \biggr)^{-3}
~{\rm days}
\end{equation}
The luminosity of the light curve, as viewed perpendicular to the
orbital plane, will be roughly that of a blackbody at $T_t$ with the
given projected surface area.
\begin{equation}
\begin{split}
L &= 4 \pi \Rs^2  \zeta^2 E_z^{2/3} E_t^{-1/3}
\sigma T_t^4  \\
&\approx 2 \times 10^{40}~\zeta_3^2 \rs^2
\biggl( \frac{E_z}{100} \biggr)^{2/3} 
\biggl( \frac{10}{E_t} \biggr)^{1/3}
\biggl( \frac{T_t}{5000~{\rm K}} \biggr)^4
~\ergss
\end{split}
\end{equation}
The disruption of solar mass stars should give rise to rather brief,
dim transients, while those of more massive stars will be significantly
brighter given their greater $\Rs$ and $\zeta_R$.

\section{Light Curves}

\begin{figure}
\includegraphics[width=3.5in]{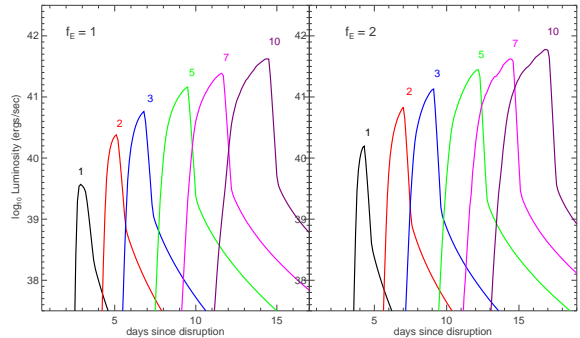}
\caption{Bolometric light curves of recombination transients from the tidal 
disruption of stars with masses $M = 1,2,3,5,10$ and $20$~\msun (from
left to right, marked on figure). The calculations assume $M_b =
10^{6}~\msun$, $\beta = 3, E_t = 10, E_z = 100$ and a viewing angle
perpendicular to the orbital plane.  The left and right panels show
calculations for two different values of the initial internal energy
(Eq.~\ref{Eq:init_energy}). The temperature and density structures of
the debris were assumed homogeneous.
\label{Fig:LC_mass}}
\end{figure}

A realistic calculation of the optical emission from tidal disruption
would require a 3-dimensional coupled radiation-hydrodynamics
simulation. Here we perform simple 1-D calculations to illustrate the
approximate luminosity and duration of the light curves.

The hydrodynamic simulations of \cite{Brassart_2008} show that shocks
tend to homogenize the debris structure in the z-direction.  We
consider models of initially uniform temperature and density $\ps =
M_{\rm ej}/V$, where the ejected mass is $M_{\rm ej} = \Ms/2$ and the
initial volume $V = 4\pi \Rs^3/3$.  The initial energy is given by
Eq.~\ref{Eq:init_energy} with $f_E = 1$ or 2.  To estimate the light
curves for stars of different masses, we apply a simple mass-radius
relation $\Rs/R_\odot = (M/M_\odot)^{0.8}$.

Radiation transport is most relevant along the $z$-direction, which we
resolve with 1000 zones.  The diffusion time is long in the ionized
central regions, so radiation transport is only of critical importance
in the surface layers.  We therefore calculate the adiabatic evolution
as described in \S\ref{sec:thermo} while including radiative losses
relevant for the cooling recombination wave, namely
\begin{equation}
\dot{E}_{\rm rad} = e^{-\bar{\tau_z}} \delta t \kappa \rho c a T^4
\end{equation}
where $\delta t$ is the duration of the timestep.  The isotropic
equivalent luminosity at a
given time, as seen from the brightest viewing angle, is
\begin{equation}
L(t) =  4 \pi \Rs^2 \zeta^2 E_z^{2/3} E_t^{-1/3}
\sigma \int d\tau e^{-\tau} T^4(\tau)
\label{Eq:RT_lum}
\end{equation}
where $\tau$ is the optical depth along the z-direction
(Eq.~\ref{Eq:tau}).  Given the small z-dimension, we neglect light
crossing times.

Figure~\ref{Fig:LC_mass} shows light curves of disruption events for
stars of masses $M = 1-10$~\msun.  We have taken a penetration factor
of $\beta = 3$ and axis ratios $E_z = 100, E_t = 10$. As expected, the
light curves rise at about 3 to 5 days after disruption, and last
around a few days.  Following the peak there is a more slowly
declining light curve tail in which the neutral, cold ($T \sim
2000$~K) debris continues to radiate in the infrared.  A proper
description of the transport in this late phase would require
inclusion of molecular opacities.

A solar mass star reaches a peak luminosity of $4\times
10^{39}$~ergs/sec for $f_E =1$ and $2 \times 10^{40}$~ergs/sec for
$f_E = 2$.  The brightnesses increases strongly with \Ms, mainly
because the more massive stars have larger radii and hence greater
surface area.  In addition, more massive stars have initially larger
internal energies and lower average densities, which allows for larger
expansion factors (\S\ref{sec:thermo}). The disruption of $\Ms
>10~\msun$ stars, though presumably rare events, would produce quite
luminous recombination transients, approaching the brightness of
Type~II supernovae ($\sim 10^{42}$~ergs/sec) although with much
shorter durations.

The emission of the recombination transient will be anisotropic due to
the asymmetry of the debris.  To first order this angular dependence
is given by the projected surface area of the photosphere along
different viewing angles.   For $E_z \gg 1$ the angular dependence is essentially $L
\propto \cos(i)$, where $i$ is the viewing angle relative to the direction perpendicular
to the orbital plane.

The spectrum of the recombination transient should be roughly
blackbody at the transparency temperature $T_t \approx 5000$~K.  The
ongoing recombination may also lead to strong emission in $H_\alpha$.
Thus the RT will likely be brightest in the R-band. The spectral
features should is some ways resemble those of Type~II supernovae,
with P-Cygni profiles.  The observed Doppler shifts, however, should
be significantly lower, corresponding to velocities $v_\infty/E_z \sim
600~\beta~\kms$.  Given the extreme asymmetry, high levels of
continuum linear polarization, near the limit of $\sim 11\%$ in plane
parallel electron scattering atmospheres \citep{Chandra_1960} should
be observed from some viewing angles.

\section{Prospects for Detection}

\subsection{Domain of Astrophysical Relevance}

\begin{figure} 
\plotone{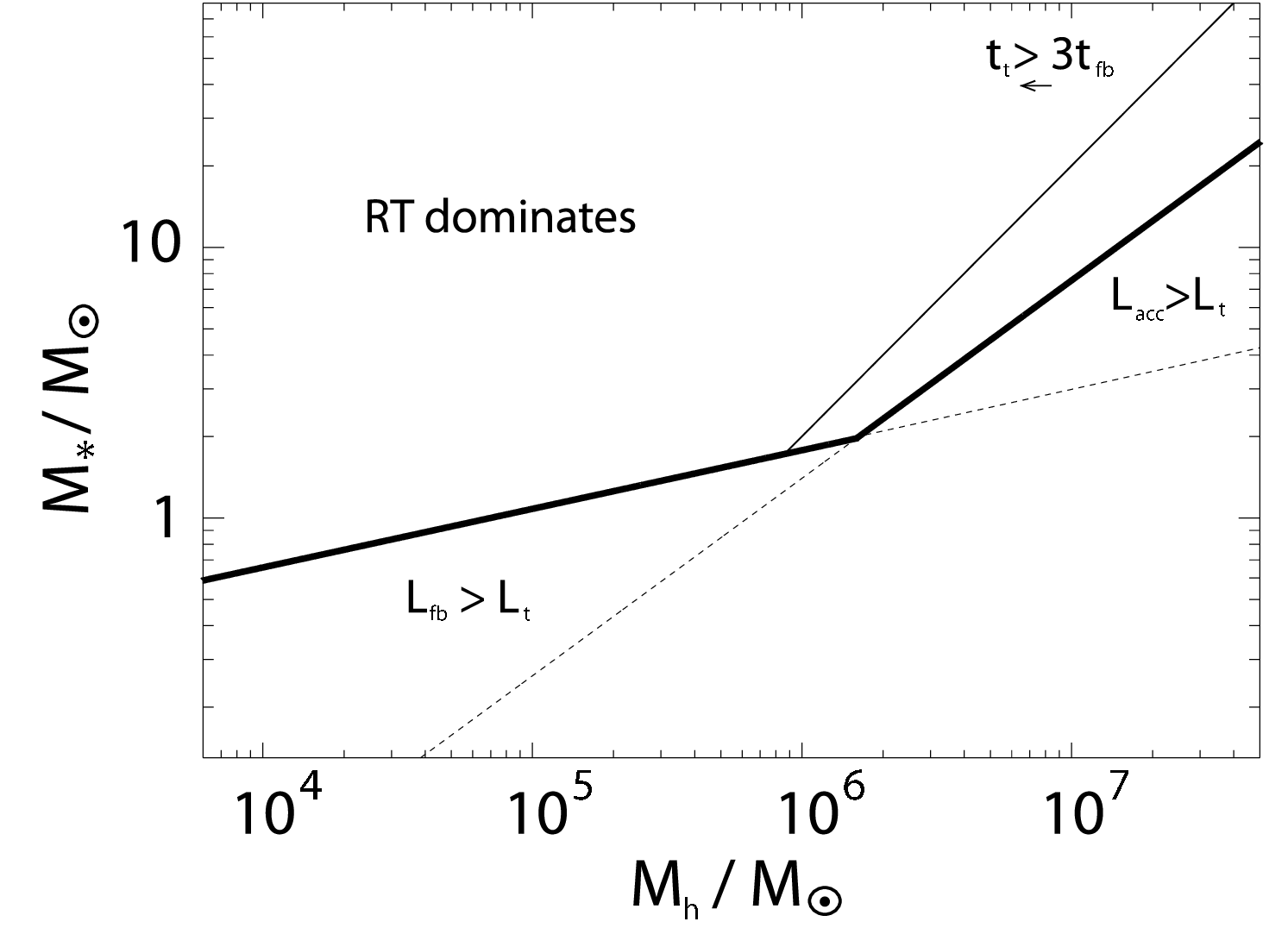}
\caption{Domain in $\Ms$-$\Mbh$
space where the emission from the unbound debris exceeds the optical luminosity
of accretion and fallback, as defined in the text.  The recombination transient
dominates in the region  above the thick solid
line (i.e., smaller black hole mass \Mbh\ and larger stellar mass \Ms).  The
thin solid line shows the region where the time of peak for the recombination transient is comparable
to the fallback timescale. \\} 
\label{asrel} 
\end{figure}  

The recombination transient is just one of several sources
of electromagnetic radiation in a tidal disruption event.  An
important question is whether the luminosity of the RT can compete
with the other emission mechanisms, in particular the fallback of the
bound stellar debris and its subsequent accretion onto the black hole.

The energy radiated during the fallback phase may dominate the
luminosity at early times.  The most tightly bound debris moves on an
elliptical orbit with semi-major axis $a = \frac{1}{2} R_p (R_p/\Rs)$
and returns to pericenter on the orbital timescale $\tfb = 2 \pi
(a^3/G \Mbh)^{1/2}$, or
\begin{equation}
\tfb \approx 40 ~\beta^{-3} m_6^{1/2} \rs^{3/2}
m_\ast^{-1}~{\rm days},
\end{equation}
with less bound material falling back continually thereafter.  The
streams of fallback material undergo collisions which, after a few
orbital timescales, circularize the orbits at a radius $\sim 2 R_p$
determined by total angular momentum conservation
\citep{Ramirez_dis09}.  The final binding energy of this circular
orbit is much less than the binding energy of the initially elliptical
orbits.  Assuming the energy difference is radiated with efficiency
$\epsilon$ over a few orbital timescales, $\sim 3\tfb$, the luminosity
is would be
\begin{equation}
\begin{split}
L_{\rm fb}  &= \frac{1}{2} \frac{G \Mbh}{2 R_p} 
 \frac{\Ms}{2}  \frac{1}{3 \tfb}  \\
 &\approx  4.6 \times 10^{43}~\beta^4~\epsilon_{0.1} m_6^{1/6} \ms^{7/3} \rs^{-5/2}\;{\rm
  erg\;s^{-1}}.
 \end{split}
 \end{equation}
where $\epsilon_{0.1} = 0.1 \epsilon$ is highly uncertain.  
Assuming the photosphere is near $2 R_p$, the emission
temperature is
\begin{equation}
 T_{\rm fb}\sim 2.4 \times 10^5~\beta^{3/2} m_6^{-1/8} \ms^{3/4
 }r_\ast^{-9/16}\;{\rm K}.
\end{equation}
It is possible that some of the energy available in fallback goes into
driving a mass outflow \citep{Strubbe_TD}.  This matter
adiabatically expands to large radii until becoming optically thin.  
For certain choices of the fraction of mass ejected, the
resulting optical luminosity can be quite bright $L \sim
10^{42}-10^{43}~\ergss$.

After circularization, the fallback material forms a torus which will
spread out viscously and accrete onto the black hole.  The timescale
for this stage of the evolution is determined by the viscous timescale
\citep{Cannizzo_1990,Ulmer_1999}
\begin{equation}
t_{\rm acc}\sim 73 
 \biggl( \frac{\alpha}{0.1} \biggr)^{-1}
 \biggl( \frac{h}{0.1} \biggr)^{-2}
\ms^{-1/2} \rs^{3/2}\;{\rm days}.
\end{equation}
where $\alpha$ is the standard viscosity parameter
\citep{shakurasun73}, and $h$ is the ratio of the disk height to
radius and is approximately of order unity for thick disks.  The
emergent radiation may have a strong dependence on the structure and
orientation of the disk. As a simple estimate we consider the
luminosity to be at the Eddington limit $L_{\rm ed} \approx 10^{44}~m_6~\ergss$ and emanating from the radius $2 R_p$.  This gives an
accretion temperature
\begin{equation}
T_{\rm acc} \sim 1.6 \times 10^5~\beta^{1/2}
m_6^{1/12} \ms^{1/6} \rs^{-1/2}\;{\rm K}.
\end{equation}
Both the fallback and the accretion temperature depend weakly on \Mbh.
In the case where emission peaks in the soft x-rays, we will observe
in the optical the Raleigh-Jeans tail of the blackbody, where the flux
is reduced by a factor $(T_{\rm fb}/T_t)^{-2}$ from its peak value.

From the above discussion, we can estimate the conditions under which
the recombination transient is a relevant source of emission.  Using
Eq.~\ref{Eq:RT_lum} and assuming a simple mass-radius relation $\rs
= \ms^{0.8}$, we find that the optical luminosity of the RT exceeds
that of fallback and accretion, respectively, when the mass of the
disrupted star exceeds
\begin{equation}
\begin{split}
\ms &\gtrsim~1.8~ (\beta/3)^{15/28} \epsilon_{0.1}^{15/28}
m_6^{25/112} \\ {\rm for}\;L_{\rm t} &\gtrsim L_{\rm fb}(T_{\rm
  fb}/5000~{\rm K})^{-2} \
 \end{split}
\end{equation}
and
\begin{equation}
\begin{split}
\ms &\gtrsim~1.4~ (\beta/3)^{-15/17} m_6^{25/34} \\ {\rm for}\;L_{\rm
  t} &\gtrsim L_{\rm acc}(T_{\rm acc}/5000~{\rm K})^{-2} \\
\end{split}
\end{equation}
Figure~\ref{asrel} depicts the relevant $\Ms$-$M_{\rm h}$ domain where
the RT is likely to contribute significantly to the optical
luminosity.  While these estimates are subject to several
uncertainties, it appears that the RT of $\ms \approx 1$ disruptions
will be most relevant when $\Mbh < 10^6~\msun$.  For more massive
black holes, disruption of larger stars may be necessary. The RT appears to be
comparable to or brighter than the optical emission resulting from the reprocessing of x-rays in the unbound debris,
studied by \cite{Strubbe_TD}.  However, if
the super-Eddington outflows also discussed by \cite{Strubbe_TD} occur, their optical luminosity can, for
certain wind parameters,  be as high as $\sim 10^{42}-10^{43}$~ergs, which 
would completely dominate the RT of even massive stars, at least at early times.  

Even if the RT luminosity falls below the other sources, it may be
discernible if it is separated in time.  From Eq.~\ref{Eq:RT_t}, the
ratio of the RT timescale, $t_t$ to the accretion/fallback timescales,
$\sim 3 \tfb$, is
\begin{equation}
{t_{\rm t} \over 3  t_{\rm fb}} \sim 0.1~\beta^2 m_6^{-2/3} \ms^{2/3}.
\end{equation}
Thus for black holes $\Mbh > 10^6~\msun$ the RT might be
observable as a precursor to the black hole accretion and fallback
signatures (Figure \ref{asrel}). In this case it may serve as a
confirming signature that the event being witnessed is in fact a tidal
disruption event.

The detection of disruption transients also requires that the optical
emission reach some significant fraction of the luminosity of the host
galaxy nucleus.  At a distance of 400 Mpc, ground based optical
surveys with a resolution of a $\sim 1$'' will only resolve a
kiloparsec-sized bulge. \cite{mclure02} give a empirical correlation
between bulge luminosity $L_b$ and central black hole mass.  For $\Mbh
= 10^8~\msun$, the bulge luminosity $L_b \approx 2 \times
10^9~L_\odot$, and only the disruption of $\ga 10$~\msun\ stars could
contribute $\ga 10\%$ of the core brightness for galaxies.  For $\Mbh
\la 10^7~\msun$, on the other hand, the disruption of $\ms>3$ stars
would outshine the galaxy and be easily detectable. However, firm
evidence for the assumed $\Mbh-L_b$ correlation below $ \Mbh <
10^7M_\odot$ is still lacking.

\subsection{Detection Rates}

Observational surveys are being designed which will probe the
magnitude range ($M_{\rm bol} = -11$ to $-14$) of the recombination
transients. LSST, for example, will reach an R-band magnitude of $M_R
= 24$~mag per shot. In this case, a $M=1~M_\odot$ disruption would be
detectable out to $\sim 100$~Mpc, the distance to the Coma cluster,
while an $M=10~M_\odot$ disruption would be detectable much further
out, to $z \approx 0.1$.

\cite{wm04} derived a total disruption rate for early-type galaxies
and bulges by combining an expression for the disruption rate
\begin{equation}
\dot{N}\approx 2.2\times 10^{-4}\;{\rm
  yr^{-1}}\;h^{-0.25}(L_g/10^{10}h^{-2}L_\odot)^{-0.295}
\end{equation}
 with \cite{fs91} E+SO luminosity function for the galaxy luminosities $L_g$. They found a rate per unit
 volume of $\sim 10^{-5}\;{\rm yr^{-1}\;Mpc^{-3}}$, with galaxies with
 $L_g \leq 10^{9}L_\odot\ (\Mbh \leq 10^8M_\odot)$ dominating the
 consumption rate. Assuming that galaxies with $L_g \ll 10^{9}L_\odot$
 hosting black holes are at least as common, we estimate that LSST may
 detect RTs of more massive stars at a rate of $\geq 300 \psi_{>5} \;{\rm yr}^{-1}$, where
 $\psi_{>5}$ is the fraction of total disruptions of stars with
 $m_\ast\geq 5$.

The major uncertainty in this rate is whether low luminosity galaxies
contain a nuclear black hole, as  M32 is currently the faintest system
($L_g \sim 4\times 10^8 L_\odot$) for which there is solid kinematical
evidence. The galaxies in question are dwarf
elliptical (dE) galaxies, which are the most numerous type of galaxy
in the Universe. Tidal disruption events are also more common in dE
galaxies, and if they contain nuclear black holes they should dominate
the total tidal flaring rate. Assuming that only nucleated dE galaxies
(dEn) contain black holes, \cite{wm04} found a total tidal disruption
for dEn in Virgo of $\sim 0.2\;{\rm yr^{-1}}$.  The overall rates in
the Coma Cluster are expected to be significantly larger.  These event
rates will be even higher if moderately massive black holes are
present in every dE galaxy or in bulges of late-type spirals.

Suggestive evidence has accumulated that intermediate mass black holes
($\Mbh = 10^3-10^4~\msun$) exist in some globular clusters
\citep[e.g.][]{noyola08}. The rate of tidal disruption events in these
systems is expected to be $\sim 10^{-7}\;{\rm yr^{-1}}$ per globular
cluster \citep{bme04}.  Taking a globular cluster space density of
$n_{\rm gc} \sim 4\;{\rm Mpc}^{-3}$ \citep{brodie06} we estimate the
density rate of RT to be at most $\sim 4 \times 10^{-7} {\rm
  yr^{-1}}\;{\rm Mpc}^{-3}$.  In this case, LSST should detect RT
events arising from the disruption of a $1~M_\odot$ star with a rate
of $\sim 5 \psi_{>1} {\rm yr^{-1}}$. The globular cluster luminosity
function in the galaxies studied to date \citep{brodie06} is well fit
by a Gaussian distribution with a peak $M_V \sim -7.4$ and $\sigma
\sim 1.4$.  The RT should thus easily outshine the cluster's optical
luminosity. Non-detection of flares after a few years would argue
against the existence of intermediate mass black holes in globular
clusters.

\section{Summary and Conclusions}

The purpose of this paper is has been to explore whether the unbound
debris of tidal disruption might give rise to luminous optical
emission, and to determine under what conditions this emission may be
observationally relevant.  The analytical results predict a brief (3-5
day) optical transient with peak luminosities in the range $2\times
10^{40}(M_\ast/M_\odot)^{5/2}$~\ergss.  The energy for the emission is stored mostly in the
ionization energy of hydrogen, and is released when the ejecta becomes
neutral and transparent at temperatures $T \approx 5000$~K.  This
physics will be relevant not only for the tidal disruption of stars by
massive black holes, but to the weak ejection of hydrogen envelopes in
general by, e.g., stellar collisions or pulsations.

The predicted bolometric luminosities of the recombination transients
fall into the interesting -- that is to say, relatively unexplored --
range between novae and supernovae.  Upcoming synoptic surveys will
begin to probe these sorts of intermediate events.  The RT of
1~\msun\ stars are relatively dim events which would only be visible
in the nearby universe ($\sim 100$~Mpc).  The disruption of more
massive stars ($M \ga 5~\msun$), on the other hand, can be quite
bright.  Such events may not be as rare as one might first expect.
There is strong evidence that the initial mass function of stars
within the Galactic center is top heavy \citep{alexander05}.  Perhaps
massive stars are more common in the cores of other galaxies as well.

For massive black holes, $\Mbh > 10^6~\msun$, the optical luminosity
from the accretion of bound material likely surpasses the emission
from the unbound debris.  Nevertheless, the RT may still be
discernible as a precursor, and may provide a important complementary
signature. While the properties of the accretion flare depends most
sensitively on the black hole mass, the luminosity of the RT depends
primarily on the properties of the disrupted star, with little
dependence on \Mbh.  In principle observations of both components
would allow one to solve for the parameters of the event, providing
insight into both the black hole mass and stellar population in the
centers of distant galaxies.  More basically, detection of a RT would
provide confirmation that the event witnessed was in fact a legitimate
tidal disruption,

Recombination transients appear to be of greatest interest for probing
lower mass black holes ($M_{\rm h} < 10^6M_\odot$). In these systems,
the optical luminosity of the transient is likely to exceed that of
accretion, and should outshine the nucleus of the galaxy. This offers
one means of searching for low mass black holes in dwarf galaxies, or
for intermediate mass black hole in globular clusters.  If black holes
do lurk in these numerous locations, the recombination transient might
be the most prominent optical signature of a stellar
disruption.

\acknowledgments We thank J. Bloom, S. Gezari, J. Guillochon,
E. Quataert, X. Prochaska, S. Rosswog, Martin Rees, L. Strubbe, and
the referee for useful discussions and comments.  Support for DK was provided 
by NASA through Hubble fellowship grant
\#HST-HF-01208.01-A awarded by the Space Telescope Science Institute,
which is operated by the Association of Universities for Research in
Astronomy, Inc., for NASA, under contract NAS 5-26555.  This research
has been supported by the DOE SciDAC Program (DE-FC02-06ER41438).
ER acknowledge support from the David and Lucile Packard
Foundation.

\end{document}